%% file: main.tex
\documentclass[lettersize,journal]{IEEEtran}
\usepackage{amsmath,amsfonts}
\usepackage{algorithmic}
\usepackage{algorithm}
\usepackage{array}
\usepackage[caption=false]{subfig}
\usepackage{textcomp}
\usepackage{stfloats}
\usepackage{url}
\usepackage{verbatim}
\usepackage{graphicx}

\usepackage[utf8]{inputenc} 
\usepackage[T1]{fontenc}    
\usepackage{url}            
\usepackage{booktabs}       
\usepackage{amsfonts}       
\usepackage{nicefrac}       
\usepackage{microtype}      
\usepackage{xcolor}         

\usepackage{makecell}
\usepackage{amsmath}
\usepackage{accents}
\usepackage{statex}
\usepackage[normalem]{ulem}
\usepackage{enumitem}
\usepackage{multirow}

\usepackage{changepage}
\usepackage{bm}
\usepackage{mathrsfs}

\usepackage{diagbox}
\usepackage{pdftexcmds}
\usepackage{catchfile}
\usepackage{ifluatex}
\usepackage{ifplatform}
\usepackage{threeparttable}

\usepackage{comment}
\usepackage{aecompl}
\usepackage{lipsum}   
\begin{document}

\title{A Unified Model for Human Mobility Generation \\in Natural Disasters}

\author{Qingyue Long, Huandong Wang, ~\IEEEmembership{Member,~IEEE}, Qi Ryan Wang, Yong Li, ~\IEEEmembership{Senior Member,~IEEE}
\thanks{Manuscript received July 20, 2025; revised xxxx, 2025.}
\thanks{The authors are with the Department of Electronic Engineering, Beijing National Research Center for Information
Science and Technology (BNRist), Tsinghua University, Beijing 100084, China
and  the Department of Civil and Environmental Engineering, Northeastern University, Boston, MA 02115.
(e-mail: longqy21@mails.tsinghua.edu.cn;
wanghuandong@tsinghua.edu.cn;
q.wang@northeastern.edu;
liyong07@tsinghua.edu.cn)
}
}

\markboth{IEEE TRANSACTIONS ON KNOWLEDGE AND DATA ENGINEERING,~VOL.~36, NO.~12, JULY~2025}%
{Shell \MakeLowercase{\textit{et al.}}: A Unified Model for Human Mobility Generation in
Natural Disasters}

\IEEEpubid{0000--0000~\copyright~2025 IEEE}

\maketitle

\input{abstract}


\input{Introduction}
\input{relatedwork}
\input{preliminaries}

\input{method}
\input{experiments}

\section{Conclusion}

In this paper, we propose a unified model named UniDisMob for human mobility generation under disaster scenarios. The model demonstrates strong generalization across different cities and disaster types. We design a physics-informed neural network to capture the underlying common patterns of mobility disruptions caused by various disasters. Moreover, UniDisMob introduces a meta-learning mechanism to address the heterogeneity among different cities and enhance adaptability.
Extensive experiments on multiple real-world datasets show the superiority of our model, particularly under zero-shot settings.
In the future, we plan to broaden the unified model's applicability to various tasks and explore the integration of more data modalities, such as social network data, to enhance human mobility modeling.

\bibliographystyle{IEEEtran}

\bibliography{sample-base}


\vfill

\end{document}

%% file: abstract.tex
\begin{abstract}
Human mobility generation in disaster scenarios plays a vital role in resource allocation, emergency response, and rescue coordination. During disasters such as wildfires and hurricanes, human mobility patterns often deviate from their normal states, which makes the task more challenging. 
However, existing works usually rely on limited data from a single city or specific disaster, significantly restricting the model’s generalization capability in new scenarios. In fact, disasters are highly sudden and unpredictable, and any city may encounter new types of disasters without prior experience. Therefore, we aim to develop a one-for-all model for mobility generation that can generalize to new disaster scenarios.
However, building a universal framework faces two key challenges: 1) the diversity of disaster types and 2) the heterogeneity among different cities. 
In this work, we propose a unified model for human mobility generation in natural disasters (named UniDisMob). To enable cross-disaster generalization, we design physics-informed prompt and physics-guided alignment that leverage the underlying common patterns in mobility changes after different disasters to guide the generation process. To achieve cross-city generalization, we introduce a meta-learning framework that extracts universal patterns across multiple cities through shared parameters and captures city-specific features via private parameters.
Extensive experiments across multiple cities and disaster scenarios demonstrate that our method significantly outperforms state-of-the-art baselines, achieving an average performance improvement exceeding 13\%.

\end{abstract}

%% file: introduction.tex
\section{Introduction}




In recent decades, the frequency and intensity of natural disasters have increased notably, encompassing a wide range of extreme events such as hurricanes, wildfires, and floods. Facing these sudden or potential disasters, human mobility generation becomes critical for evacuation route planning~\cite{xu2023interconnectedness}, rescue resource allocation~\cite{tang2023resilience}, and long-term social reconstruction~\cite{zhang2022human}. However, disasters may induce mobility perturbations, breaking people’s routine travel patterns, as reflected in shifts in travel distances, changes in regular routes, and even evacuations to temporary shelters~\cite{li2022spatiotemporal}. These complex and dynamic mobility patterns have also posed significant challenges to accurately generating human mobility under disaster scenarios. 

By leveraging advanced generative AI techniques, numerous studies have focused on human mobility generation~\cite{long2025one, jin2023time, zhu2023difftraj}. However, the majority of these studies primarily focus on capturing the periodic patterns of human mobility under normal scenarios, which limits their applicability in disaster scenarios~\cite{long2023practical, feng2020learning, wang2023pategail}. 
Therefore, a few studies have addressed mobility modeling in disaster scenarios. For example, DeepMob leverages deep neural networks to learn human mobility patterns during earthquakes from multi-source data of millions of users, enabling accurate trajectory prediction~\cite{song2017deepmob}. MemeSTN combines memory networks and meta-learning to integrate social media data with human mobility data, enabling human mobility nowcasting in disaster scenarios~\cite{jiang2023learning}.
Although these studies have made significant progress in modeling human mobility under disaster scenarios, their effectiveness is often confined to specific disaster types within a single city. Moreover, their reliance on large-scale training data from diverse cities and disaster types further limits the flexibility and generalizability of such models.

\begin{figure}[t]
\centering
\includegraphics[width=0.47\textwidth]{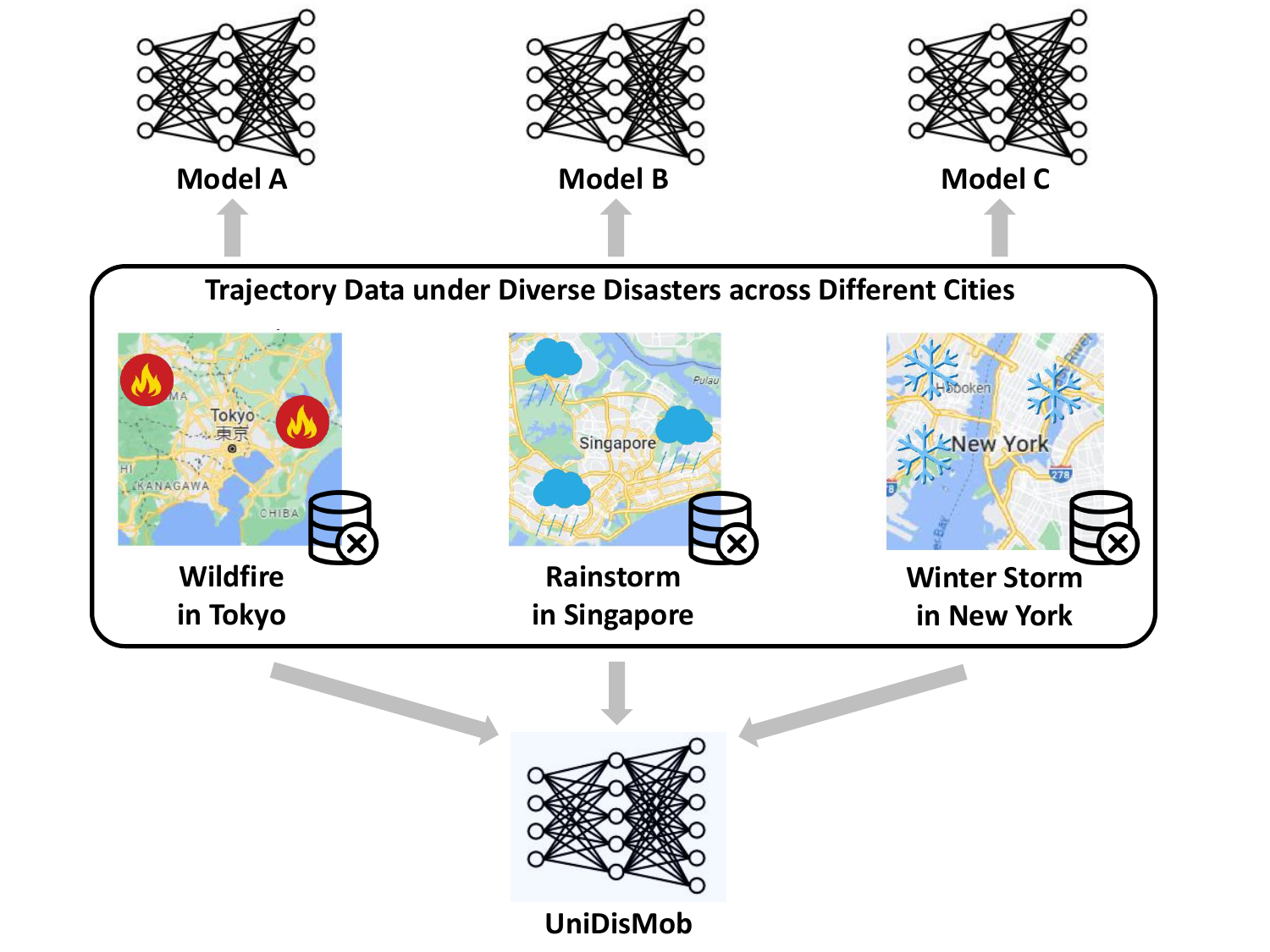}
\caption{The transition from single model to universal model.}
\label{fig:UniDisMob}
\vspace{-0.3cm}
\end{figure}

In fact, disasters are highly sudden and unpredictable, and any city may encounter new types of disasters without prior experience. If models are trained exclusively on the limited trajectory data collected under specific disaster scenarios in individual cities, they often struggle to generalize to new cities and novel disaster types~\cite{liu2021towards, lakara2021evaluating}. As shown in Figure~\ref{fig:UniDisMob}, we aim to develop a unified model for human mobility generation in natural disasters that captures shared spatiotemporal patterns across multiple cities and disasters.
However, a unified model for human mobility generation in
natural disasters is still an open problem with the following challenges:
\begin{itemize}[leftmargin=*]
\item \textbf {Diversity of disaster types.} Different types of disasters exhibit significant differences in disturbance intensity, duration, and spatial extent. For example, hurricanes may trigger prolonged large-scale evacuations, whereas earthquakes often result in short-term but high-intensity changes in mobility patterns. Therefore, it’s challenging to model the impact of different disaster types on human mobility within a unified model. 

\item \textbf {Heterogeneity across different cities.} Differences across various cities in population density, spatial layout, and other factors lead to complex and diverse human mobility patterns. For example, when the target city and the source city differ significantly in geographical location and spatial structure, the resulting location embedding features are often substantially different. This heterogeneity makes it difficult to achieve cross-city transfer with a unified model.
\end{itemize}

\IEEEpubidadjcol

To address the above challenges, we propose a unified model for human mobility generation in natural disasters named \textbf{UniDisMob}.
Firstly, we design a physics-informed neural network to achieve effective generalization across multiple disaster types. Specifically, we introduce physics-informed prompts that leverage spatiotemporal decay mechanisms derived from physical models~\cite{li2022spatiotemporal} to capture the latent spatiotemporal patterns of mobility changes following different disasters, which serve as prior knowledge to guide trajectory generation. Moreover, we construct a physics-guided alignment mechanism. The alignment mechanism is implemented via a physics-guided loss, which enforces consistency between the generated mobility patterns and the expected spatiotemporal decay. This loss function incorporates constraints based on known spatiotemporal decay laws of mobility under disaster scenarios into the training objective, which enhances the interpretability and robustness of the unified model.
Secondly, we adopt a meta-learning framework to address the heterogeneity across different cities. This framework extracts common knowledge from multiple cities while leveraging adaptation mechanisms to fine-tune the model for specific cities. In this way, it balances generalization and flexibility. Specifically, the framework consists of shared and private parameters. The shared parameters are responsible for learning spatiotemporal patterns that are consistent across cities and extracting knowledge applicable to various disaster scenarios. In contrast, the private parameters capture city-specific mobility characteristics, enabling the model to adapt to diverse urban structures.
Therefore, UniDisMob achieves strong generalization across cities and disaster types by learning the common patterns of mobility changes under different disaster scenarios. Overall, our contributions can be summarized as follows:
\begin{itemize}[leftmargin=*]
\item We explore the potential of a one-for-all model for mobility generation in natural disasters, achieving generalization across different cities and disaster types.

\item We propose a unified model for human mobility generation in natural disasters. Specifically, we design a physics-informed neural network for cross-disaster generalization and a meta-learning framework to extract shared knowledge across cities for cross-city generalization.

\item We conduct extensive experiments on seven datasets covering different cities and disaster types. The results demonstrate that our method outperforms state-of-the-art models in mobility generation, achieving an average performance improvement exceeding 13\%. Further analysis confirms UniDisMob’s strong zero-shot capability, with an average performance improvement of over 8\% on new cities.
\end{itemize}

%% file: relatedwork.tex
\section{Related Work}
\subsection{Spatial-Temporal Foundation Models}
In recent years, inspired by foundation models, particularly large language models, researchers have begun to explore spatio-temporal foundation models (STFMs) to enhance adaptability and generalization across a wide range of spatio-temporal tasks~\cite{xue2024prompt, gruver2023large, jin2023large}. Existing approaches can be broadly categorized into two main types:
The first category is LLM-based methods, which transform spatio-temporal data into text-like sequences and leverage the universal representation capabilities of LLMs to achieve cross-domain transfer and few-shot learning~\cite{liu2024lstprompt, zhong2025time, li2024streetviewllm}. For example, Mai et al.~\cite{mai2023opportunities} investigate how large pretrained language models can be utilized for representation learning on geo-spatial multimodal data, thereby improving remote sensing analysis and geographic information reasoning. Similarly, Time-LLM~\cite{jin2023time} reprograms raw time series into structured textual prototype representations, thereby directly leveraging the powerful representation and reasoning capabilities of LLMs for prediction and analysis. UrbanGPT~\cite{li2024urbangpt} transforms complex urban spatio-temporal data into structured textual inputs and leverages instruction tuning to build a large language model for spatio-temporal tasks. 
The second category is a pretrained foundation model, which is pretrained directly on large-scale, multi-domain spatio-temporal data to build model architectures with efficient spatio-temporal representation capabilities~\cite{cao2024timedit, wen2023diffstg, yuan2024unist}. For example, UrbanDiT~\cite{yuan2024urbandit} is designed for urban spatio-temporal tasks, providing a unified model that integrates diverse spatio-temporal data sources and types while learning general spatio-temporal patterns across different cities and scenarios. UniST~\cite{yuan2024unist} is a universal spatio-temporal foundation model that leverages large-scale pre-training and prompt adaptation to extract general spatio-temporal patterns across different domains. 
Although foundation models have made some progress in the spatiotemporal domain, there is still a significant gap in universal mobility modeling under disaster scenarios. In this work, we propose UniDisMob, the first attempt to build a universal model for mobility generation in disaster scenarios.

\subsection{Human Mobility Generation}
Human mobility generation can be divided into trajectory generation under normal scenarios and trajectory generation under disaster scenarios. 
Trajectory generation under normal scenarios focuses on people’s travel patterns in daily life, capturing the periodicity and regularity of their mobility behaviors. Rule-based methods, such as EPR~\cite{song2010modelling} and TimeGeo~\cite{jiang2016timegeo}, model the spatio-temporal regularities of human mobility behaviors to generate trajectories. Deep generative models learn the distribution of trajectory data to generate human mobility trajectories. For example, PateGail~\cite{wang2023pategail} leverages an adversarial learning framework to learn individual behavior policies from trajectories, generating high-fidelity mobility patterns. DiffTraj~\cite{zhu2023difftraj} employs a diffusion-based denoising process to model the complex spatiotemporal distribution of trajectories, enhancing the diversity of the generated outcomes.
Different from trajectory generation under normal scenarios, trajectory generation under disaster scenarios focuses more on modeling human mobility behaviors that deviate from normal patterns during disasters. For example, Liang et al.~\cite{li2022spatiotemporal} propose a spatial-temporal decay mechanism, which models how mobility perturbations after a disaster gradually diminish over time and with increasing spatial distance, offering a new perspective on mobility generation. Song et al.~\cite{song2014prediction} develop an emergency mobility simulator based on large-scale disaster data to generate high-fidelity trajectories under disaster scenarios. Subsequently, Song et al.~\cite{song2017deepmob} design a framework integrating behavioral models and spatio-temporal dependencies to model human mobility patterns following natural disasters.
However, these methods are typically trained for a specific disaster type or city and lack generalization across diverse scenarios. In this work, we develop a universal model with cross-city and cross-disaster generalization capability for trajectory generation under disaster scenarios.

%% file: preliminaries.tex
\section{Preliminaries}

\subsection{Problem Definition}
\textit{Definition 1: (Mobility Trajectory).} The mobility trajectory of a user $u$ is defined as an ordered sequence of locations recorded at uniform time intervals, denoted as $x_u = {l_1, l_2, \dots, l_n}$. Each location $l_i$ corresponds to either a geographic coordinate (latitude and longitude) or a designated region identifier.

\textit{Definition 2: (Disaster Intensity).} $N = \left\{ N_j(t)\mid j \in \mathcal{L},\ t \in \mathcal{T} \right\}$
 denotes the disaster intensity, where $N_j(t)$ represents the disaster intensity at location $j$ and time $t$. $\mathcal{L}$ denotes the set of all locations, and $\mathcal{T}$ denotes the set of all time slots.

\textit{Problem Definition: (Human Mobility Generation Problem in Disaster Scenario).}
Given disaster intensity $N$ and a set of real-world mobility trajectories $X = \{x_u\}$, the goal is to learn a generative model that can generate mobility trajectories while preserving the fidelity of real-world data.

\subsection{Conditional Denoising Diffusion Probabilistic Models}
\label{sec:PRELIMINARIES}
Diffusion models are a class of latent variable models characterized by the expression $p_\theta(x_0) := \int p_\theta(x_{0:T}), dx_{1:T}$, where the latent variables $x_1, \dots, x_T$ share the same dimensionality as the observed data $x_0 \sim q(x_0)$. These models employ two Markov chains: a forward process that gradually adds noise to the data, and a reverse process that aims to recover the original input. The following Markov chain defines the forward (diffusion) process:
\begin{equation}\label{equ:DDPM1}
q\left(\mathbf{x}_{1: T} \mid \mathbf{x}_{0}\right):=\prod_{t=1}^{T} q\left(\mathbf{x}_{t} \mid \mathbf{x}_{t-1}\right), 
\end{equation}
where $q\left(\mathbf{x}_{t} \mid \mathbf{x}_{t-1}\right):=\mathcal{N}\left(\sqrt{1-\beta_{t}} \mathbf{x}_{t-1}, \beta_{t} \mathbf{I}\right)$ and $\beta_t$ are modest positive constants that reflect the noise intensity. $x_t$ may be represented in closed form as $x_{t}=\sqrt{\alpha_{t}} x_{0}+\left(1-\alpha_{t}\right) \epsilon.$ for $\epsilon \sim \mathcal{N}(0, \mathbf{I})$, where $\alpha_{t}=\sum_{i=1}^{t}\left(1-\beta_{t}\right)$.

We turn to the conditional extension~\cite{tashiro2021csdi}, where the target variable $x$ is generated given a set of observed conditions $c$. The objective of the conditional denoising diffusion probabilistic model is to approximate the conditional distribution $q(x_0|c)$ using a parameterized model $p_\theta(x_0|c)$.

In contrast, the conditional reverse process is defined by the following Markov chain, which iteratively denoises $x_t$ in order to recover $x_0$.
\begin{equation}\label{equ:DDPM2}
p_{\theta}\left(\mathbf{x}_{0: T}|c\right):=p\left(\mathbf{x}_{T}\right) \prod_{t=1}^{T} p_{\theta}\left(\mathbf{x}_{t-1} \mid \mathbf{x}_{t},c\right),
\end{equation}
where $\mathbf{x}_{T} \sim \mathcal{N}(\mathbf{0}, \mathbf{I})$. The conditional transition distribution $p_\theta(x_{t-1} \mid x_t, c)$ is modeled as a Gaussian with learnable parameters, given by:
\begin{equation}\label{equ:DDPM3}
p_{\theta}\left(\mathbf{x}_{t-1} \mid \mathbf{x}_{t},c\right):=\mathcal{N}\left(\mathbf{x}_{t-1} ; \boldsymbol{\mu}_{\theta}\left(\mathbf{x}_{t}, t|c\right), \sigma_{\theta}\left(\mathbf{x}_{t}, t|c\right) \mathbf{I}\right).
\end{equation}
As shown by Ho et al.~\cite{ho2020denoising}, the unconditional denoising diffusion model adopts a specific parameterization for the transition distribution $p_\theta(x_{t-1} \mid x_t)$, formulated as follows:
\begin{equation}\label{equ:DDPM4}
\begin{cases}
\boldsymbol{\mu}_{\theta}\left(\mathbf{x}_{t}, t\right)=\frac{1}{\alpha_{t}}\left(\mathbf{x}_{t}-\frac{\beta_{t}}{\sqrt{1-\alpha_{t}}} \boldsymbol{\epsilon}_{\theta}\left(\mathbf{x}_{t}, t\right)\right), \\
\sigma_{\theta}\left(\mathbf{x}_{t}, t\right)=\tilde{\beta}_{t}^{1 / 2} \quad \text{where} \quad \tilde{\beta}_{t}=\left\{
\begin{array}{ll}
\frac{1-\alpha_{t-1}}{1-\alpha_{t}} \beta_{t} & t>1 \\
\beta_{1} & t=1
\end{array}\right.
\end{cases}
\end{equation}
where $\epsilon_\theta$ is a trainable denoising function. In Eq.~\ref{equ:DDPM4}, we represent $\boldsymbol{\mu}_{\theta}\left(\mathbf{x}_{t}, t\right)$ and $\sigma_{\theta}\left(\mathbf{x}_{t}, t\right)$ as $\mu^{\mathrm{DDPM}}\left(\mathbf{x}_{t}, t, \boldsymbol{\epsilon}_{\theta}\left(\mathbf{x}_{t}, t\right)\right)$ and $\sigma^{\mathrm{DDPM}}\left(\mathbf{x}_{t}, t\right)$.

Building on the unconditional diffusion model, a conditional denoising function $\epsilon_\theta$ is introduced, which incorporates the condition $c$ as an additional input. We now analyze the resulting parameterization involving $\epsilon_\theta$:
\begin{equation}\label{equ:DDPM5}
\begin{cases}
\boldsymbol{\mu}_{\theta}\left(\mathbf{x}_{t}, t \mid c\right) = \boldsymbol{\mu}^{\mathrm{DDPM}}\left(\mathbf{x}_{t}, t, \boldsymbol{\epsilon}_{\theta}\left(\mathbf{x}_{t}, t \mid c\right)\right), \\
\sigma_{\theta}\left(\mathbf{x}_{t}, t \mid c\right) = \sigma^{\mathrm{DDPM}}\left(\mathbf{x}_{t}, t\right).
\end{cases}
\end{equation}

To optimize the denoising function $\epsilon_\theta$, the model minimizes the following loss function:
\begin{equation}\label{equ:DDPM6}
\min _{\theta} \mathcal{L}(\theta):=\min _{\theta} \mathbb{E}_{\mathbf{x}_{0} \sim q\left(\mathbf{x}_{0}\right), \boldsymbol{\epsilon} \sim \mathcal{N}(\mathbf{0}, \mathbf{I}), t}||\boldsymbol{\epsilon}-\boldsymbol{\epsilon}_{\theta}\left(\mathbf{x}_{t}, t|c\right)||_{2}^{2},
\end{equation}
where $\mathbf{x}{t} = \sqrt{\alpha{t}} \mathbf{x}{0} + (1 - \alpha{t}) \boldsymbol{\epsilon}$. The denoising function $\epsilon_\theta$ aims to predict the noise vector $\boldsymbol{\epsilon}$ added to the perturbed input $\mathbf{x}_t$. This formulation reduces the influence of tokens at very small $t$ (i.e., low noise levels), and can be interpreted as a variational constraint weighted by a negative log-likelihood.

%% file: method.tex
\section{Method}
\begin{figure*}[t]
\centering
\includegraphics[width=0.95\textwidth]{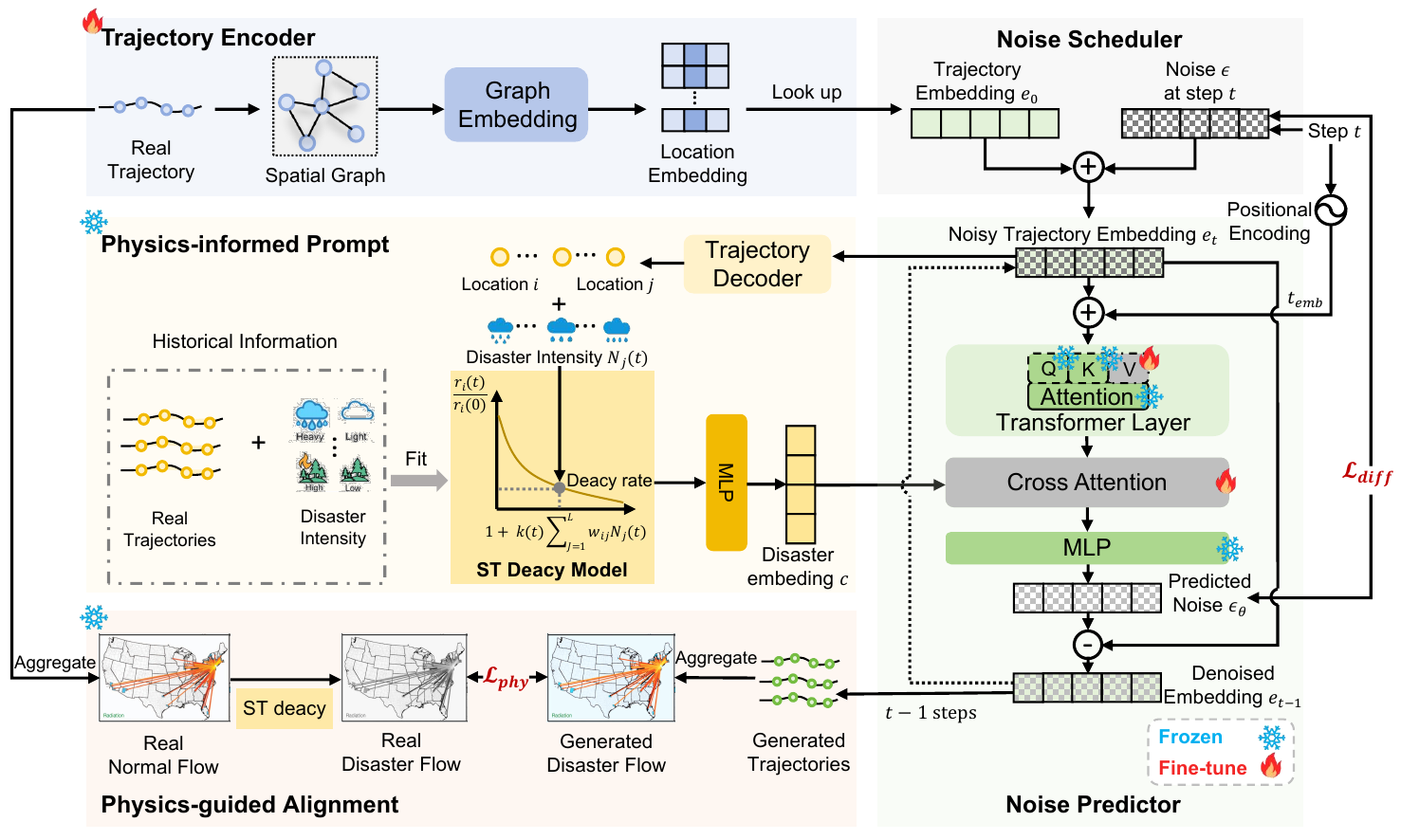}
\caption{The overview architecture of UniDisMob, which consists of five modules: (1) Trajectory Encoder, (2) Noise Scheduler, (3) Physics-informed Prompt, (4) Noise Predictor, (5) Physics-guided Alignment. } \label{fig:framework}
\vspace{-0.3cm}
\end{figure*}

As shown in Figure~\ref{fig:framework}, the overview architecture of UniDisMob consists of five modules: trajectory encoder, noise scheduler, physics-informed prompt, noise predictor, and physics-guided alignment. Among them, the trajectory encoder, noise scheduler, and noise predictor together serve as common components for diffusion-based trajectory generation.
To achieve cross-disaster generalization, we design a physics-informed neural network that incorporates physical priors of human mobility dynamics after disasters to improve adaptability to new disaster scenarios. 
Specifically, we first introduce physics-informed prompts that leverage a spatiotemporal decay model to capture the shared patterns of mobility changes following different disasters, serving as conditions to guide the trajectory generation process. 
Subsequently, we employ a physics-guided alignment mechanism that integrates the spatiotemporal decay patterns under disasters into the training objectives, aligning the generated trajectories with the physical patterns to ensure more realistic outcomes.
To address the heterogeneity across different cities, we adopt a meta-learning framework that extracts shared knowledge from multiple cities and quickly adapts to specific cities through an adaptation mechanism.

\begin{figure*}[t]
\centering
\includegraphics[width=0.95\textwidth]{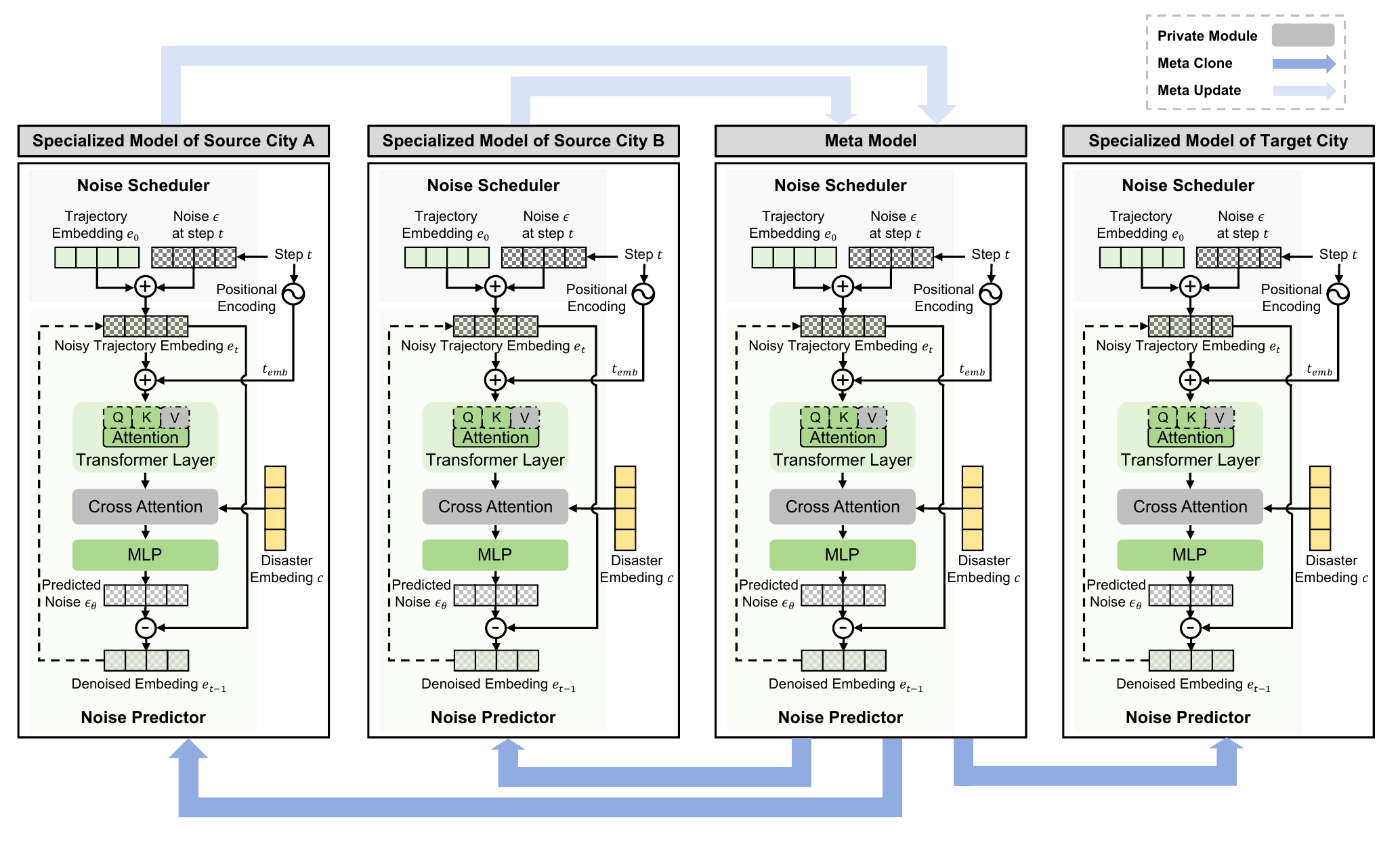}
\caption{The meta-learning framework, which extracts common patterns across multiple cities through shared modules, while capturing city-specific characteristics via private modules.}
\label{fig:diffusion}
\vspace{-0.3cm}
\end{figure*}

\subsection{Trajectory Encoder}
Following the common strategies adopted in previous works~\cite{yuan2023spatio,zhou2023towards}, we perform diffusion and denoising processes on the trajectory embedding.
Firstly, we construct two types of temporal embedding matrices: $P \in \mathbb{R}^{T_d \times w}$ and $Z \in \mathbb{R}^{T_h \times W}$, to characterize the temporal dependencies and periodic patterns of mobility behaviors. Specifically, the matrix $P$ encodes the weekly periodic patterns, where $T_d = 7$ denotes the number of days in a week. The matrix $Z$ represents the temporal features of different time intervals within a day (e.g., morning and evening peaks), where $T_h$ denotes the number of time slots determined by the sampling granularity. Through temporal embeddings, the model can effectively capture the periodicity and regularity in the temporal dimension.

Secondly, to capture the spatial characteristics of mobility behaviors, we construct a spatial graph $G = (V, E)$ to model the geographical continuity of trajectory points, ensuring that each transition spans a limited spatial range. In this graph, the node set $V$ represents all visited locations, while the edge set $E$ defines the topological relationships between locations, reflecting their spatial proximity and potential interactions. Each edge $e = (u, v)$ is an unordered pair associated with a positive weight $w_{uv}$, which quantifies the Euclidean distance between locations $u$ and $v$, thereby characterizing the spatial closeness among locations.
We adopt a graph embedding function $\mathcal{F}(\cdot)$ to derive the location embedding $D$, which is denoted as follows:
\begin{equation}\label{equ:graph}
D={\mathcal{F}}_\theta(G),
\end{equation}
This process yields a spatial embedding matrix $D \in \mathbb{R}^{L \times W}$, where $L$ denotes the total number of regions and $W$ indicates the embedding dimension.

Finally, we concatenate the spatial and temporal embeddings to form the comprehensive trajectory representation, as follows:
\begin{equation}\label{equ:traj_embedding}
e=[D; P; Z],
\end{equation}
where $e$ denotes the embedding vector of the trajectory, integrating both spatial and temporal features to provide high-dimensional, information-rich inputs for generation tasks.

\subsection{Noise Scheduler}
To enhance the robustness of trajectory representation learning and promote the generation of high-quality samples, we adopt a noise scheduling mechanism that progressively injects noise into the trajectory embeddings. Specifically, given an initial clean embedding $e_0$, the noise injection at diffusion step $t$ is implemented by adding Gaussian noise to the embedding, formally defined as:
\begin{equation}\label{equ:noisy}
    \mathbf{e}_t = \sqrt{\bar{\alpha}_t} \mathbf{e}_0 + \sqrt{1 - \bar{\alpha}_t} \boldsymbol{\epsilon},
\end{equation}
where $\epsilon \sim \mathcal{N}(0, \mathbf{I})$ denotes standard Gaussian noise, and $\bar{\alpha}_t = \prod_{s=1}^{t} \alpha_s$represents the cumulative product of noise scaling factors. The sequence $\{\alpha_s\}$ controls the scheduling strategy for noise intensity, which can be smoothly adjusted using linear or cosine annealing.

\subsection{Physics-informed Prompt}
\label{sec:Physics-informed Prompt}
The extent and duration of mobility changes caused by different disasters vary significantly across disaster types. Therefore, it is necessary to rely on explicit physical interpretations to model the unified patterns of mobility changes. The spatiotemporal decay model (ST decay model) captures the latent spatiotemporal patterns of crowd mobility following various disasters and reveals the underlying consistencies within these changes~\cite{li2022spatiotemporal}.

The spatiotemporal decay model demonstrates that across various types of disasters, despite differences in their causes and durations, the resulting disturbances in human mobility consistently exhibit a unified hyperbolic decay pattern. On the one hand, there is a spatial decay effect: the closer a population is to the disaster’s core area, the more pronounced the reduction in mobility, while those farther away experience relatively minor impacts. On the other hand, there is a temporal decay effect: as time progresses, the demand to resume normal activities gradually increases, leading to a recovery in mobility. Based on these observations, this study proposes a hyperbolic model that jointly captures spatiotemporal decay dynamics as follows:

\begin{equation}\label{equ:st_deacy}
    r_i(t) = \frac{r_i(0)}{1 + k(t) \sum_{j=1}^{L} w_{ij} N_j(t)},
\end{equation}
where $w_{ij} N_j(t)$ characterizes spatial decay, $w_{ij}$ represents the spatial weight between location $i$ and $j$, and $N_j(t)$ denotes the intensity of the disaster in location $j$ at time $t$.

$k(t)$ characterizes temporal decay. As time progresses, the initial changes in mobility behavior gradually diminish until they eventually disappear. Therefore, $k(t)$ is a time-dependent dynamic function, expressed by the following equation:
\begin{equation}\label{equ:temporal_deacy}
    k(t) = k_0 e^{-\alpha t},
\end{equation}
where $\alpha$ is the parameter that controls the rate of decay. $k(0)$ represents the initial rate of change in mobility behavior, which is assumed to reach its maximum at $t = 0$. As $t$ becomes sufficiently large, $k(t)$ gradually approaches zero.

Based on historical trajectories collected under various types of disasters and their corresponding disaster intensities, we fit the aforementioned spatiotemporal decay model to characterize the underlying patterns across different disasters, as formulated below:
\begin{equation}\label{equ:st}
    H_\theta(i,t) = \frac{r_i(t)} {r_i(0)} = \frac{1}{1 + k(t) \sum_{j=1}^{L} w_{ij} N_j(t)},
\end{equation}

At each step of the denoising process, the model uses the current noisy trajectory embedding $e_t$ and decodes it into a sequence of locations through the trajectory decoder:
\begin{equation}\label{equ:decoder}
s_t = \{l_1,...l_i\} = \arg\max \,\mathrm{sim}\bigl(\mathbf{e}_t,\, \mathbf{D}\bigr),
\end{equation}
where $sim(\cdot)$ is a similarity function, and $\arg\max$ returns the index corresponding to the maximum value.

The decoded location $i$, the current time $t$, and the corresponding disaster intensity $N$ are input into the fitted ST decay model $H_\theta(i,t)$ to obtain the mobility decay rate $r_i$ for each location. Then, we use an MLP to obtain the disaster embedding $c$, which models the representation of the impact of the current disaster conditions on mobility. The formulation is as follows:
\begin{equation}\label{equ:disaster_emb}
c_i = \text{MLP}(r_i),
\end{equation}
In the denoising network, the disaster embedding $c$ serves as a prompt that provides contextual information about the impact of disasters on mobility. Thus, it dynamically guides the generation process and ensures that the generated trajectories conform to disaster-specific patterns.

\subsection{Noise Predictor}
At each step of the denoising process, the model leverages the denoised embedding $\epsilon_t$, the current diffusion step $t$, and the disaster embedding $d$ to achieve condition-guided denoising. The noise predictor takes these inputs to output the predicted noise:
\begin{equation}\label{equ:predicted noise}
\hat{\epsilon}_{\theta}=\epsilon_{\theta}({{e}_t}, t, c).
\end{equation}

To obtain the representation of the diffusion step $t$, we follow previous works~\cite{vaswani2017attention, kong2020diffwave} and adopt a positional encoding approach, encoding $t$ into a 128-dimensional vector $t_{emb}$ as follows:
\begin{equation}\label{equ:diffusion step}
\begin{aligned}
t_{emb} = & \left[\sin \left(10^{0 \times 4 / 63} t\right), \ldots, \sin \left(10^{63 \times 4 / 63} t\right), \right. \\
& \left.\cos \left(10^{0 \times 4 / 63} t\right), \ldots, \cos \left(10^{63 \times 4 / 63} t\right)\right].
\end{aligned}
\end{equation}

We add the positional encoding of the diffusion step $t$ to the trajectory embedding $e_t$ at step $t$:
\begin{equation}
z_t = e_t + t_{emb}.
\end{equation}

Then, we utilize a transformer to capture the spatio-temporal dependencies of trajectories, formulated as:
\begin{equation}
h_t = \text{Transformer}(z_t).
\end{equation}

To introduce dynamic guidance from disaster conditions during the trajectory denoising process, we fuse the disaster embedding $d$ obtained from the physics-informed prompt module with the trajectory representation $h_t$ using a cross-attention mechanism. Specifically, the disaster embedding $d$ is used as the key and value input, while the current trajectory representation $h_t$ serves as the query. The computation is as follows:
\begin{align}
\tilde{h}_t &= \text{CrossAttn}(h_t W_Q, d W_K, d W_V) \notag \\
            &= \text{softmax} \left( \frac{h_t W_Q (d W_K)^\top}{\sqrt{d_k}} \right) (d W_V),
\end{align}
where $h_t$ denotes the trajectory features obtained from the Transformer, $d$ is the disaster embedding, $d_k$ represents the dimensionality of the key vectors, and $W_Q$, $W_K$, and $W_V$ are learnable projection matrices.

This conditional fusion mechanism allows each trajectory point to adaptively acquire relevant information from the disaster context according to its state, thus realizing the perception of the disaster context. The final output $\tilde{h}_t$ is fed into an MLP to predict the noise at the current diffusion step $t$:
\begin{equation}
\hat{\epsilon}_{\theta} = \text{MLP}(\tilde{h}_t),
\end{equation}
The output $\hat{\epsilon}_{\theta}$ is of the same shape as $e_t$. 

\subsection{Physics-guided Alignment}
To enhance the realism of trajectory generation under disaster scenarios, we introduce the physics-guided alignment module. This module aligns the generated disaster-induced human flow with the disaster flow modeled by the physical method, guiding the trajectory generation process to follow the macroscopic mobility trends influenced by disasters.

We aggregate the real trajectories under normal scenarios $X_{normal}$ and the generated trajectories $X_{gen}$ under disaster scenarios into crowd flow $F_{normal}$ and $F_{gen}$, expressed as:
\begin{align}
F_{{normal}} &= {\text{Aggregate}}(X_{{normal}}), \\
F_{{gen}} &= {\text{Aggregate}}(X_{{gen}}),
\end{align}
where $F \in \mathbb{R}^{L \times T}$, $T$ denotes the number of time slots, and $L$ represents the number of spatial locations.

We use the spatiotemporal decay model $H_\theta(i,t)$ fitted in Section~\ref{sec:Physics-informed Prompt} to estimate the disaster-induced crowd flow $F_{dis}$ based on the normal crowd flow $F_{normal}$:
\begin{equation}
F_{dis} = H_\theta(i,t)F_{normal}.
\end{equation}

We align the crowd flow $F_{gen}$, obtained by aggregating the trajectories generated by UniDisMob, with the disaster-induced crowd flow $F_{dis}$ derived from the spatiotemporal decay model, and define the physics-guided loss function $\mathcal{L}_{\text{phy}}$ as follows:
\begin{equation}
\mathcal{L}_{phy} = || F_{gen} - F_{dis} ||_2^2.
\end{equation}

This loss serves as an additional supervisory signal during training, constraining the generated trajectories to align with the decay patterns observed under disaster scenarios at the population level.

\subsection{Training and Sampling}
\paragraph{Training}
As shown in Figure.~\ref{fig:diffusion}, to address the heterogeneity across different cities, we adopt a meta-learning framework that extracts universal mobility patterns from multiple source cities and enables quick adaptation to each target city. The modules marked as frozen and fine-tuned in Figure.~\ref{fig:framework} reflect the division between shared and private components. The shared modules learn cross-city common mobility patterns for knowledge transfer, while the private modules capture city-specific features to enhance local adaptability.

Our loss function consists of the diffusion model loss and an additional physics-guided loss. Based on the derivation in Section~\ref{sec:PRELIMINARIES}, the diffusion model is trained using a simplified loss function:
\begin{equation}\label{equ:diffusion}
\mathcal{L}_{diff} = \mathbb{E}_{{e}_0,{\epsilon}, t}||\boldsymbol{\epsilon}-\boldsymbol{\epsilon}_{\theta}\left(\mathbf{e}_t, t|d\right)||_{2}^{2}.
\end{equation}

Therefore, the overall loss function is defined as follows:
\begin{equation}
\mathcal{L}_{total} = \alpha\mathcal{L}_{diff} + \beta\mathcal{L}_{phy},
\end{equation}
where $\alpha$ and $\beta$ are the weighting coefficients for the two loss terms. The total loss $\mathcal{L}_{total}$ encourages the model to preserve individual trajectory characteristics while generating mobility patterns that align with physical laws under disaster scenarios.

\textbf{Meta-training on Source Cities:} We extract universal mobility patterns from multiple source cities. For each source city $D^k = (D^k_{train},D^k_{val})$, we first clone the shared modules from the meta model and combine them with the city-specific private modules to form the initial model parameters:
\begin{equation}
\Theta_{src}^k = \Theta_{shared}^{meta} \cup\Theta_{private}^k.
\end{equation}

Next, internal update is performed on the training set $D^k_{train}$ of the city to minimize the total loss $\mathcal{L}_{total}$ and update both the shared and private modules:
\begin{equation}
\Theta_{src}^k \leftarrow \Theta_{src}^k - \alpha_{src} \nabla_{\Theta} \left( L_{total} \right).
\end{equation}

Then, the updated model is evaluated on the validation set $D_{val}^k$ of the corresponding city, and the gradients are computed to update only the shared modules (Meta Update):
\begin{equation}
\Theta_{meta} \leftarrow \Theta_{meta} - \alpha_{meta} \nabla_{\Theta_{shared}} \left( L_{total} \right).
\end{equation}

\textbf{Target Adaptation:} After completing meta-training on the source cities, we select the target city dataset $\tilde{D} = \left( \tilde{D}_{train}, \tilde{D}_{test}\right)$, clone the shared module parameters, and combine them with the private modules of the target city to obtain the initialized model parameters:
\begin{equation}
\Theta_{tgt} = \Theta_{shared}^{meta} \cup \Theta_{private}^{tgt}.
\end{equation}

Then, fine-tuning is performed on the training set $\tilde{D}_{train}$ of the target city, aiming to minimize the total loss function:
\begin{equation}
\Theta_{tgt} \leftarrow \Theta_{tgt} - \alpha_{tgt} \nabla_{\Theta} \left( L_{total} \right).
\end{equation}

Finally, the model is evaluated on the test set $\tilde{D}_{test}$ of the target city and used to generate trajectories under disaster scenarios.

\paragraph{Sampling}
We sample using a linear combination of conditional and unconditional predictions~\cite{ho2022classifier}:
\begin{equation}\label{equ:denoising}
{\tilde{\epsilon}_{\theta}} = (1+\omega){\epsilon}_{\theta}({{e}_{t}}, t|d) - \omega{\epsilon}_{\theta}({{e}_{t}}, t|\varnothing),
\end{equation}
where $\varnothing$ indicates a vector that effectively does not encode any information. When $\omega=0$, the model is fully conditional; $\omega=1$ yields an unconditional model. Increasing $\omega$ balances diversity and sample quality.

Denoising proceeds iteratively from step $T$ to 1. At each step, noise is removed via:
\begin{equation}\label{equ:sampling}
e_{t-1}={\frac{1}{\sqrt{\alpha_{t}}}(e_{t}-\frac{1-\alpha_{t}}{\sqrt{1-\bar{\alpha_{t}}}}\tilde{\epsilon}_{\theta})}+{\sigma_t}z
\end{equation}
where $\tilde{\epsilon}_{\theta}$ estimates the noise $\epsilon$ from from $e_{t}$, and $z \sim N (0, I)$.

%% file: experiments.tex
\section{Experiments}
\subsection{Experimental Settings}
To evaluate the performance of UniDisMob, we conduct extensive experiments on seven city datasets. A unified model is trained using data from four cities (Houston, Guilin, Phoenix, and Boston), and zero-shot testing is performed on the remaining three cities (Los Angeles, Worcester, and Sacramento). In this setting, the model generates trajectories for target cities without having seen any trajectory data under disaster scenarios, thereby assessing its cross-city generalization and transferability.

\subsubsection{Dataset}
The detailed information of the datasets is shown in Table~\ref{table:datasets}. We use seven real-world mobility datasets from different cities under various disaster types. We use four datasets (Houston, Guilin, Phoenix, and Boston) to jointly train our unified model, and conduct transfer experiments on three cities (Los Angeles, Worcester, and Sacramento) to evaluate the zero-shot capability of our model.
For data preprocessing, we filter out users with fewer than five records per day. In the spatial dimension, raw GPS coordinates are converted to predefined grid IDs at a specified granularity. In the temporal dimension, timestamps are uniformly segmented into fixed-length intervals, using half-hour intervals as the unit.

\begin{figure}[t]
\centering
\subfloat [Guilin]{\includegraphics[width=.22\textwidth]{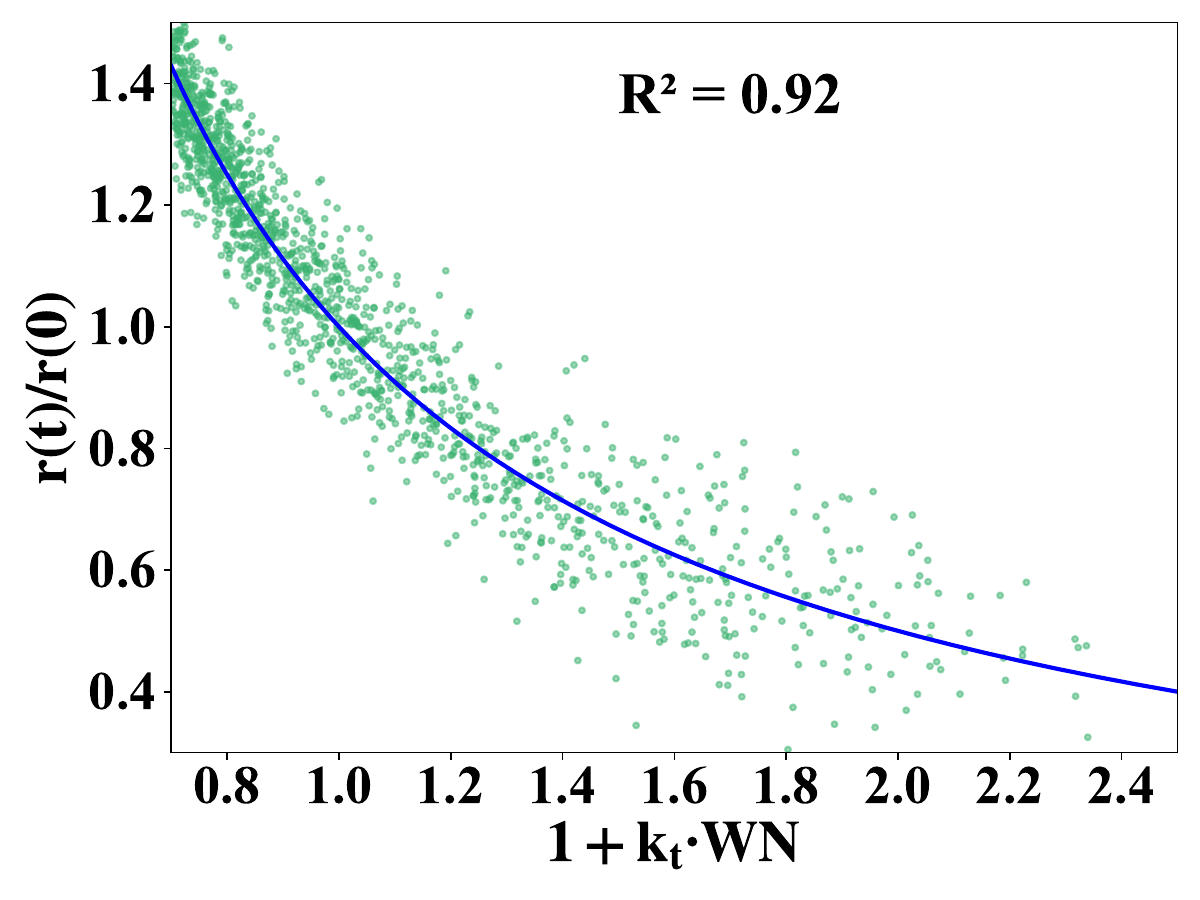}}
\hfil
\subfloat [Boston]{\includegraphics[width=.22\textwidth]{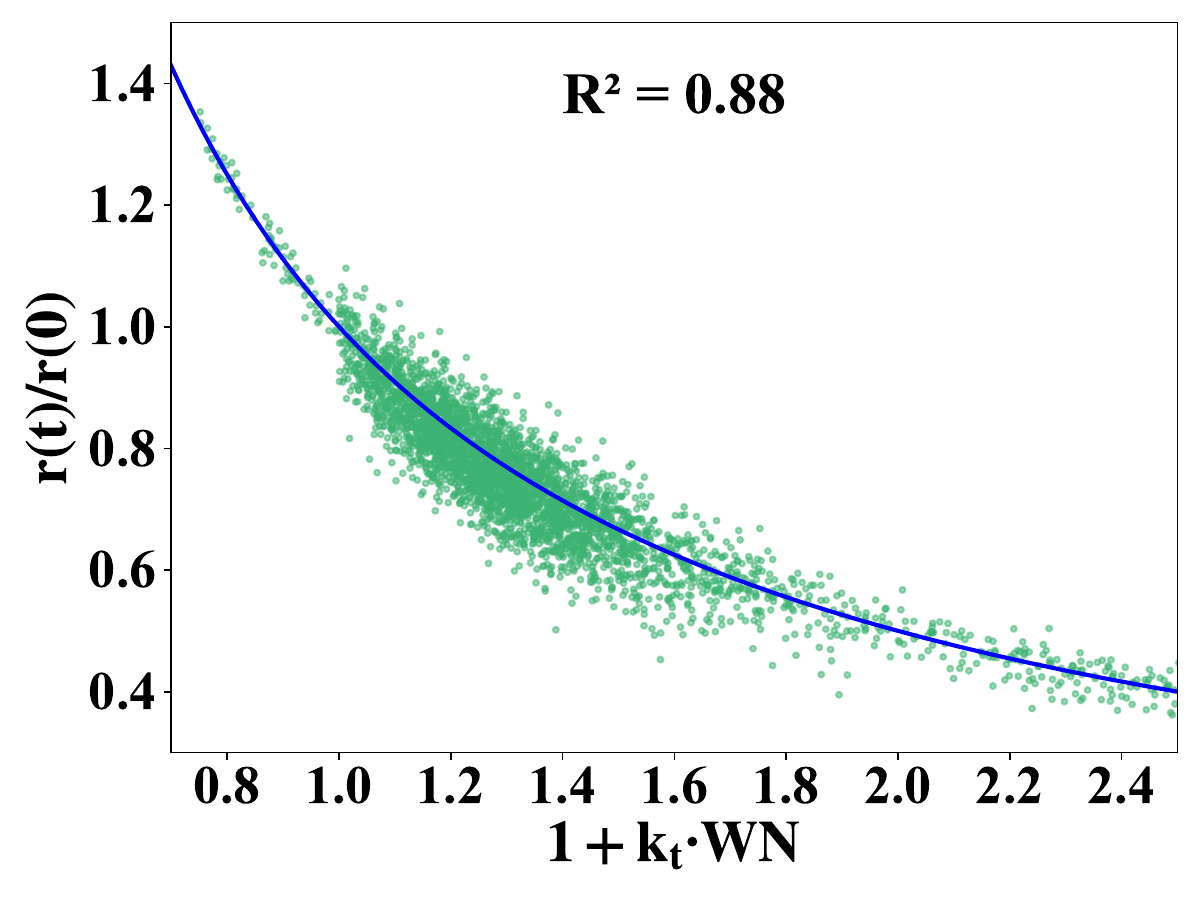}}
\caption{Fitting performance of the spatiotemporal decay model across different cities.}
\label{fig:fit}
\end{figure}

As shown in Figure~\ref{fig:fit}, we present the fitting performance of the spatiotemporal decay model, which was trained on multiple datasets, across different cities. The results indicate that the model effectively captures the decay patterns of human mobility over time and space under disaster scenarios, demonstrating its generalization ability across different types of disasters.

\begin{table*}[htbp]
\centering
\caption{Basic statistics of mobility datasets.}
\begin{tabular}{lcccccc}
\toprule
\textbf{City} & \textbf{Disaster Type} & \textbf{Data Duration} & \textbf{Disaster Duration} & \textbf{Disaster Intensity} & \textbf{\#Users} \\
\midrule
Houston       & Hurricane      & 30 days & 9 days  & Rainfall           & 24162 \\
Guilin        & Rainstorm      & 17 days & 6 days  & Rainfall           & 12035 \\
Phoenix       & Extreme Heat   & 50 days & 9 days  & Temperature Rise   & 8637  \\
Boston        & Winter Storm   & 28 days & 8 days  & Temperature Drop   & 20191 \\
Los Angeles   & Wildfire       & 40 days & 7 days  & PM2.5              & 15408 \\
Worcester     & Winter Storm   & 31 days & 7 days  & Temperature Drop   & 6810  \\
Sacramento    & Extreme Heat   & 45 days & 10 days & Temperature Rise   & 3784  \\
\bottomrule
\end{tabular}
\label{table:datasets}
\end{table*}

\subsubsection{Metrics}
We evaluate the generated trajectory using the following six metrics: The first four assess the behavioral characteristics of trajectories, and the last two measure mobility changes under disaster scenarios.

\begin{itemize}[leftmargin=*]
\item \textbf{Distance}: It refers to the total length of a user's movement paths accumulated over a specific period and is evaluated using JSD.
\item \textbf{Radius}: Radius of gyration measures the spatial extent of a user's daily activity range and is evaluated using JSD.
\item \textbf{Duration}: It indicates the time a user spends at each visited location and is evaluated using JSD.
\item \textbf{Daily-loc}: It represents the number of distinct places a user visits daily and is evaluated using JSD.
\item \textbf{Decay Rate}: The degree of mobility decline compared to the normal state after a disaster is measured using MAPE.
\item \textbf{Decay Speed}: The speed of recovery from the mobility disruption is represented by the parameter $\alpha$ in the function $k(t) = k_0 e^{-\alpha t}$, and is evaluated using MAPE.
\end{itemize}

\textbf{MAPE} is the mean absolute percentage error between the predicted and ground truth values. \textbf{JSD} is the Jensen-Shannon divergence, which is used to evaluate the similarity between the feature distributions of real trajectories and generated trajectories. The calculation formula is:
\begin{equation}\label{equ:JSD}
{\rm JSD}(\boldsymbol{a},\boldsymbol{b})=\frac{1}{2}{\rm KL}(\boldsymbol{a}||\frac{\boldsymbol{a}+\boldsymbol{b}}{2})+\frac{1}{2}{\rm KL}(\boldsymbol{b}||\frac{\boldsymbol{a}+\boldsymbol{b}}{2}),
\end{equation}
where ${\rm KL}(\cdot||\cdot)$ is the Shannon information, $a$ and $b$ are distributions.

\subsubsection{Baselines}
\paragraph{Trajectory Generation under Normal Scenario} These models are designed for human mobility generation in normal scenarios.
\begin{itemize}[leftmargin=*]
\item \textbf{TimeGeo}~\cite{jiang2016timegeo}: This method captures the temporal characteristics (travel rhythm, dwell rate, and burstiness) and the spatial characteristics (exploration and preferential return mechanisms) to generate individual trajectories.
\item \textbf{MoveSim}~\cite{feng2020learning}: This method is a GAN-based framework that leverages urban structure priors and mobility regularity-aware loss to generate more realistic trajectories.
\item \textbf{PateGail}~\cite{wang2023pategail}: PateGail is a novel method that combines GAIL with differential privacy to generate high-quality trajectories while protecting user privacy.
\item \textbf{DiffTraj}~\cite{zhu2023difftraj}: DiffTraj is a trajectory generation method based on a diffusion model, which is designed to simulate high-fidelity human mobility behavior.
\end{itemize}

\paragraph{Trajectory Generation under Disaster Scenario} These models are designed for human mobility modeling in disaster scenarios.
\begin{itemize}[leftmargin=*]
\item \textbf{DeepMob}~\cite{song2017deepmob}: DeepMob leverages deep neural networks to extract rich representations of human mobility from large-scale data and models individual movement decision-making processes under emergency scenarios.
\item \textbf{EmeMob}~\cite{song2014prediction}: This method captures key patterns of human daily mobility and models how factors such as social relationships, disaster severity, government shelters, and news reports influence mobility behavior during disasters.
\end{itemize}

\paragraph{Controllable Generation} These baselines are controllable generation models that can use disaster information as conditions to guide the trajectory generation.
\begin{itemize}[leftmargin=*]
\item \textbf{CSDI}~\cite{song2014prediction}: This method is a conditional diffusion model that can reconstruct complete data by progressive denoising under the condition of observed values.
\item \textbf{ControlTraj}~\cite{zhu2024controltraj}: This method is a controllable trajectory generation framework that leverages diffusion models combined with conditional constraints to generate high-fidelity trajectories.
\end{itemize}

\paragraph{Cross-city Generation} These baselines leverage knowledge from source cities to assist trajectory generation in target cities.
\begin{itemize}[leftmargin=*]
\item \textbf{COLA}~\cite{wang2024cola}: COLA is a cross-city human trajectory simulation framework that leverages model-agnostic meta-learning strategies to achieve knowledge transfer.
\item \textbf{CHAML}~\cite{chen2021curriculum}: This method is a cross-city prediction framework designed to leverage the experience of data-rich source cities to improve prediction performance in target cities.
\end{itemize}

\subsection{Overall Performance}

\begin{table*}[htbp]
\centering
\caption{Overall performance on Guilin and Boston datasets.}
\scalebox{0.90}{
\begin{tabular}{lcccccccccccc}
\toprule
{\textbf{Dataset}} & \multicolumn{6}{c}{\textbf{Guilin}} & \multicolumn{6}{c}{\textbf{Boston}} \\
\cmidrule(lr){2-7} \cmidrule(lr){8-13}
{Metrics↓} & Distance  & Radius  & Duration  & Daily-loc & Decay Rate & Decay Speed & Distance  & Radius  & Duration  & Daily-loc & Decay Rate & Decay Speed \\
\midrule
TimeGEO       & 0.0628 & 0.4703 & 0.0912 & 0.4257 & 28.6\% & 23.2\% & 0.0438 & 0.3614 & 0.0795 & 0.3872 & 27.8\% & 22.3\% \\
PateGail      & 0.0207 & 0.3361 & 0.0519 & 0.1841 & 20.6\% & 18.3\% & 0.0185 & 0.3149 & 0.0642 & 0.2256 & 19.7\% & 17.4\% \\
MoveSim       & 0.0236 & 0.3875 & 0.0603 & 0.2740 & 22.4\% & 19.1\% & 0.0219 & 0.3017 & 0.0594 & 0.1908 & 21.6\% & 18.9\% \\
DiffTraj      & 0.0156 & 0.2962 & 0.0480 & 0.1037 & 15.3\% & 12.8\% & 0.0152 & 0.2651 & 0.0487 & 0.1202 & 14.3\% & 12.2\% \\
EmeMob        &0.0149 &0.2826 &0.0458 &0.0915 & 13.7\% & 11.8\% &0.0151 &0.2538 &0.0471 &0.1073 & 13.2\% & 11.3\% \\
DeepMob       &0.0142 &0.2714 &0.0439 &0.0832 & 12.4\% & 10.4\% &0.0146 &0.2460 &0.0453 &0.0975 & 11.9\% & 9.7\% \\
CSDI          & 0.0135 & 0.2517 & 0.0412 & 0.0694 & 11.6\% & 9.3\% & 0.0149 & 0.2385 & 0.0465 & 0.0864 & 10.7\% & 8.8\% \\
ControlTraj   & \underline{0.0119} & \underline{0.2405} & \underline{0.0381} & \underline{0.0673} & 10.2\% & 8.2\% & \underline{0.0138} & \underline{0.2261} & \underline{0.0448} & 0.0805 & 9.6\% & 7.7\% \\
CHAML         &0.0137 &0.2612 &0.0418 &0.0723 &10.7\% &9.1\% &0.0150 &0.2351 &0.0459 &0.0815 &10.2\% &8.5\% \\
COLA          &0.0132 &0.2526 &0.0407 &0.0690 &\underline{9.3\%} &\underline{7.8\%} &0.0154 &0.2298 &0.0451 &\underline{0.0772} &\underline{8.8\%} &\underline{7.2\%} \\
Ours          &\textbf{0.0083} &\textbf{0.2059} &\textbf{0.0320} &\textbf{0.0638} &\textbf{7.8\%} &\textbf{6.2\%} &\textbf{0.0126} &\textbf{0.1842} &\textbf{0.0435} &\textbf{0.0714} &\textbf{7.4\%} &\textbf{5.9\%} \\
\midrule
Improvement &30.2\% &14.4\% &15.9\% &5.2\% &15.9\% &18.1\% &2.3\% &18.5\% &2.9\% &7.5\% &16.1\% &20.5\% \\
\bottomrule
\end{tabular}
}
\label{tab:guilin_boston}
\end{table*}

\begin{table*}[htbp]
\centering
\caption{Overall performance on Phoenix and Houston datasets.}
\scalebox{0.90}{
\begin{tabular}{lcccccccccccc}
\toprule
{\textbf{Dataset}} & \multicolumn{6}{c}{\textbf{Phoenix}} & \multicolumn{6}{c}{\textbf{Houston}} \\
\cmidrule(lr){2-7} \cmidrule(lr){8-13}
{Metrics↓} & Distance  & Radius  & Duration  & Daily-loc & Decay Rate & Decay Speed & Distance  & Radius  & Duration  & Daily-loc & Decay Rate & Decay Speed \\
\midrule
TimeGEO       & 0.0735 & 0.5184 & 0.0983 & 0.4526 & 32.4\% & 30.1\% & 0.0647 & 0.4861 & 0.1127 & 0.4076 & 30.7\% & 29.2\% \\
PateGail      & 0.0418 & 0.3897 & 0.0773 & 0.2537 & 25.3\% & 22.8\% & 0.0375 & 0.3564 & 0.0895 & 0.2283 & 23.6\% & 22.1\% \\
MoveSim       & 0.0462 & 0.4228 & 0.0835 & 0.2981 & 27.0\% & 24.1\% & 0.0451 & 0.3952 & 0.0943 & 0.2661 & 25.2\% & 23.3\% \\
DiffTraj      & 0.0357 & 0.3426 & 0.0721 & 0.1934 & 18.2\% & 16.0\% & 0.0312 & 0.3117 & 0.0855 & 0.1826 & 17.4\% & 15.6\% \\
EmeMob        &0.0331 &0.3290 &0.0692 &0.1785 & 16.5\% & 14.7\% &0.0304 &0.2989 &0.0837 &0.1704 & 15.9\% & 14.2\% \\
DeepMob       &0.0316 &0.3175 &0.0674 &0.1692 & 14.9\% & 12.9\% &0.0291 &0.2865 &0.0818 &0.1612 & 14.1\% & 12.4\% \\
CSDI          &0.0305 &0.3194 &0.0688 &0.1659 & 13.6\% & 11.6\% &0.0287 &0.2311 &0.0831 &0.1635 & 13.0\% & 11.0\% \\
ControlTraj   &0.0281 &0.3078 &0.0675 &0.1513 & 11.3\% & 9.8\% &0.0240 & \underline{0.2180} & \underline{0.0801} & \textbf{0.1285} & 11.0\% & 9.2\% \\
CHAML         &0.0270 &0.2945 &0.0660 &0.1492 &12.0\% &10.5\% &0.0235 &0.2350 &0.0830 &0.1462 &11.7\% &10.0\% \\
COLA          & \underline{0.0276} & \underline{0.3052} & \underline{0.0661} & \underline{0.1485} &\underline{10.4\%} &\underline{9.0\%} & \underline{0.0225} &0.2226 &0.0817 &0.1440 &\underline{10.1\%} &\underline{8.6\%} \\
Ours          & \textbf{0.0247} & \textbf{0.2873} & \textbf{0.0624} & \textbf{0.1251} & \textbf{8.3\%} & \textbf{7.5\%} & \textbf{0.0201} & \textbf{0.1937} & \textbf{0.0795} & \underline{0.1363} & \textbf{8.0\%} & \textbf{7.3\%} \\
\midrule
Improvement &10.5\% &5.9\% &7.5\% &17.3\% &20.2\% &16.7\% &10.6\% &11.1\% &0.8\% &- &20.8\% &15.1\% \\
\bottomrule
\end{tabular}
}
\label{tab:phoenix_houston}
\end{table*}

We evaluate the performance of various baselines and our approach across multiple mobility metrics in four cities affected by different types of disasters. We conduct multiple experiments and reported the average performance. Our method achieves an average improvement of over 13\% across the four datasets. As shown in Table~\ref{tab:guilin_boston}, our approach outperforms all baseline methods across all six metrics on both the Guilin and Boston datasets. In particular, it shows notable advantages in modeling mobility changes caused by disasters. Specifically, our method achieves an average improvement of over 18\% on the Decay Rate and Decay Speed metrics. As shown in Table~\ref{tab:phoenix_houston}, our method also demonstrates strong performance on most metrics in the Phoenix and Houston datasets. These results indicate that our method excels in modeling human mobility across spatial and temporal dimensions and in capturing mobility perturbations caused by disasters. It more accurately replicates real-world human movement patterns under disaster scenarios, especially in quantifying the impact severity and recovery dynamics.

Notably, the two best-performing baselines, ControlTraj and COLA, represent two mainstream approaches. ControlTraj is based on diffusion models and adopts a condition-controlled generation strategy, while COLA is built on a Transformer architecture and incorporates meta-learning to enable cross-city transfer. Both methods enhance generalization by leveraging either conditional generation or transfer learning techniques. Our proposed UniDisMob integrates both strengths. On the one hand, it employs a diffusion transformer guided by physics-informed priors to guide the trajectory generation process, ensuring that the generated mobility patterns align with real-world disaster-induced changes. On the other hand, it utilizes a meta-learning framework to extract universal mobility patterns across cities while adapting to city-specific characteristics, thereby achieving stronger transferability. As the first unified framework tailored for trajectory generation in disaster scenarios, UniDisMob significantly expands the frontier of research in human mobility modeling.

\subsection{Zero-shot Performance}
\begin{figure}[t]
\centering
\includegraphics[width=1.0\linewidth]{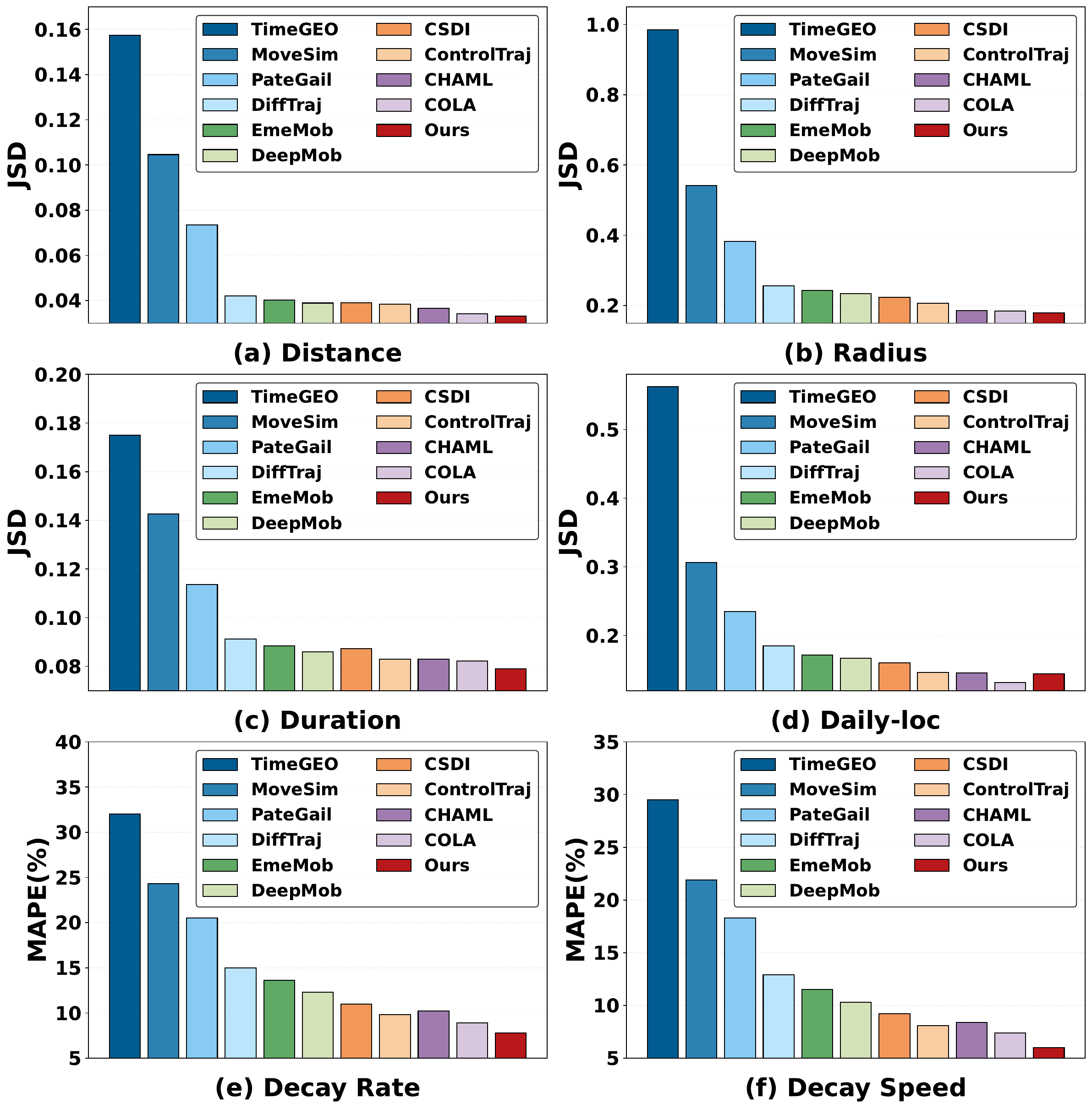}
\caption{Zero-shot performance on Los Angeles datasets.} 
\label{fig:Los Angeles}
\end{figure}

\begin{figure}[t]
\centering
\includegraphics[width=1.0\linewidth]{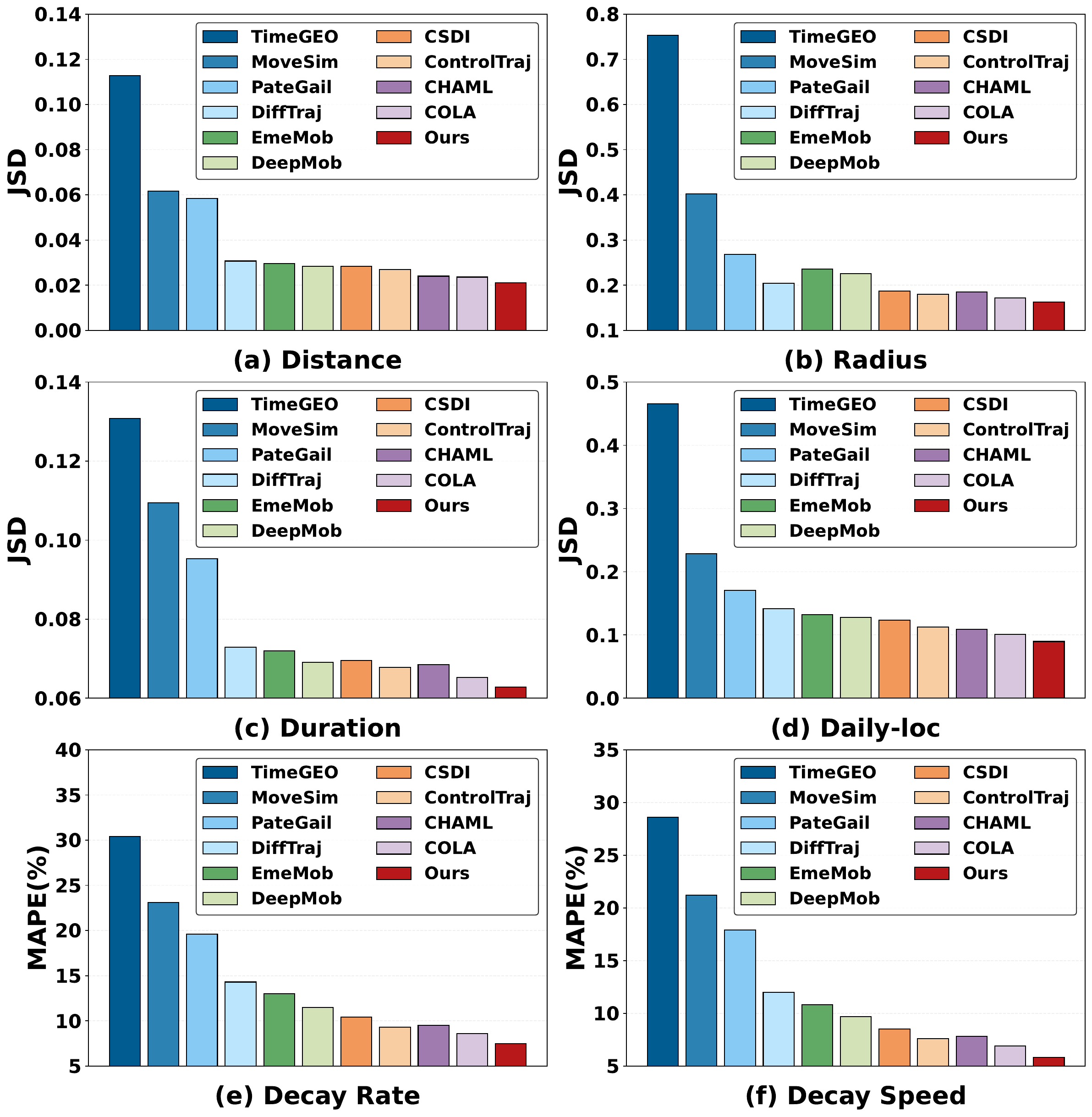}
\caption{Zero-shot performance on Worcester datasets.} 
\label{fig:Worcester}
\end{figure}

\begin{figure}[t]
\centering
\includegraphics[width=1.0\linewidth]{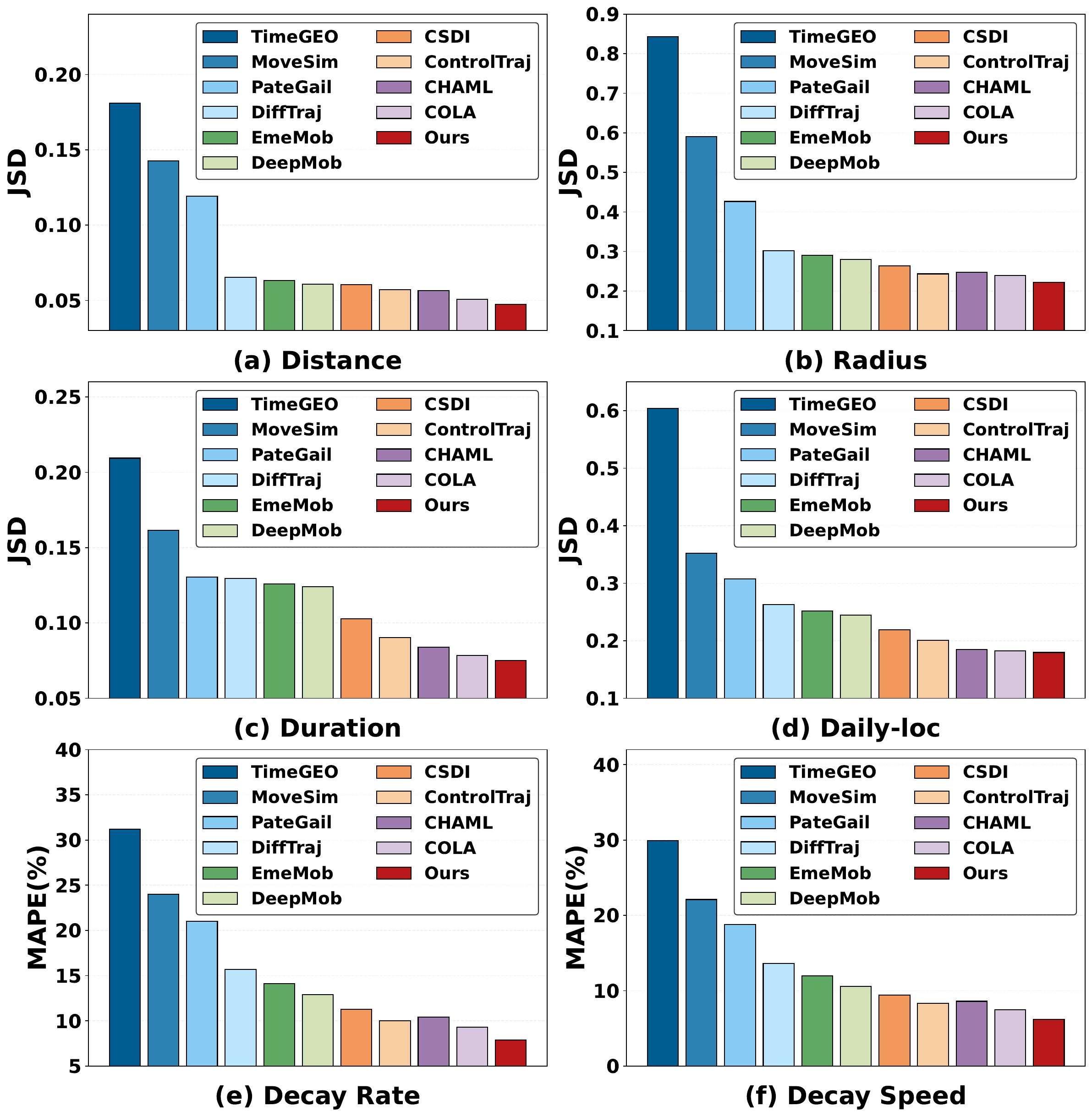}
\caption{Zero-shot performance on Sacramento datasets.} 
\label{fig:Sacramento}
\end{figure}

Figures~\ref{fig:Los Angeles}, ~\ref{fig:Worcester} and~\ref{fig:Sacramento} compare the transferability of UniDisMob with several baseline models across three representative cities (Los Angeles, Worcester, and Sacramento), each corresponding to a different type of disaster: wildfire, winter storm, and extreme heat. The transfer setting assumes that no trajectory data under disaster scenarios is available for the target city. Instead, only a small amount of trajectories under normal scenarios is provided. We leverage these normal trajectories to perform lightweight fine-tuning of the general model, enabling effective adaptation to disaster scenarios in the target city.
As shown in the figures, UniDisMob consistently achieves the best zero-shot performance across all metrics and datasets, outperforming even those baselines specifically designed for cross-city generalization. On average, it delivers over 8\% improvement across the three datasets. These results demonstrate that UniDisMob, through its physics-informed design, can effectively capture the universal patterns of mobility changes under disaster scenarios. This enables it to generalize well to entirely new cities or disaster types, showcasing strong adaptability and broad applicability.

\subsection{Ablation Study}

\begin{table*}[htbp]
\centering
\caption{Ablation results on Guilin and Boston datasets.}
\scalebox{0.85}{
\begin{tabular}{lcccccccccccc}
\toprule
\textbf{Dataset} & \multicolumn{6}{c}{\textbf{Guilin}} & \multicolumn{6}{c}{\textbf{Boston}} \\
\cmidrule(lr){2-7} \cmidrule(lr){8-13}
Metrics & Distance & Radius & Duration & Daily-loc & Decay Rate & Decay Speed & Distance & Radius & Duration & Daily-loc & Decay Rate & Decay Speed \\
\midrule
Ours & 0.0083 & 0.2059 & 0.0320 & 0.0638 & 7.8\% & 6.2\% & 0.0126 & 0.1842 & 0.0435 & 0.0714 & 7.4\% & 5.9\% \\
w/o prompt & 0.0095 & 0.2203 & 0.0351 & 0.0694 & 8.5\% & 6.8\% & 0.0136 & 0.1974 & 0.0482 & 0.0780 & 8.2\% & 6.6\% \\
\scriptsize{} & \scriptsize{(-14.5\%)} & \scriptsize{(-7.0\%)} & \scriptsize{(-9.7\%)} & \scriptsize{(-8.8\%)} & \scriptsize{(-9.0\%)} & \scriptsize{(-9.7\%)} & \scriptsize{(-7.9\%)} & \scriptsize{(-7.2\%)} & \scriptsize{(-10.8\%)} & \scriptsize{(-9.2\%)} & \scriptsize{(-10.8\%)} & \scriptsize{(-11.9\%)} \\
w/o alignment & 0.0102 & 0.2347 & 0.0376 & 0.0719 & 8.8\% & 7.1\% & 0.0145 & 0.2029 & 0.0501 & 0.0813 & 8.4\% & 6.8\% \\
\scriptsize{} & \scriptsize{(-22.9\%)} & \scriptsize{(-14.0\%)} & \scriptsize{(-17.5\%)} & \scriptsize{(-12.7\%)} & \scriptsize{(-12.8\%)} & \scriptsize{(-14.5\%)} & \scriptsize{(-15.1\%)} & \scriptsize{(-10.2\%)} & \scriptsize{(-15.2\%)} & \scriptsize{(-13.9\%)} & \scriptsize{(-13.5\%)} & \scriptsize{(-15.3\%)} \\
w/o meta learning & 0.0090 & 0.2138 & 0.0336 & 0.0659 & 7.9\% & 6.4\% & 0.0130 & 0.1895 & 0.0457 & 0.0745 & 7.6\% & 6.1\% \\
\scriptsize{} & \scriptsize{(-8.4\%)} & \scriptsize{(-3.8\%)} & \scriptsize{(-5.0\%)} & \scriptsize{(-3.3\%)} & \scriptsize{(-1.3\%)} & \scriptsize{(-3.2\%)} & \scriptsize{(-3.2\%)} & \scriptsize{(-2.9\%)} & \scriptsize{(-5.1\%)} & \scriptsize{(-4.3\%)} & \scriptsize{(-2.7\%)} & \scriptsize{(-3.4\%)} \\
\bottomrule
\end{tabular}}
\label{tab:ablation_guilin_boston}
\end{table*}

\begin{table*}[htbp]
\centering
\caption{Ablation study on Phoenix and Houston datasets.}
\scalebox{0.85}{
\begin{tabular}{lcccccccccccc}
\toprule
\textbf{Dataset} & \multicolumn{6}{c}{\textbf{Phoenix}} & \multicolumn{6}{c}{\textbf{Houston}} \\
\cmidrule(lr){2-7} \cmidrule(lr){8-13}
Metrics & Distance & Radius & Duration & Daily-loc & Decay Rate & Decay Speed & Distance & Radius & Duration & Daily-loc & Decay Rate & Decay Speed \\
\midrule
Ours & 0.0247 & 0.2873 & 0.0624 & 0.1251 & 8.3\% & 7.5\% & 0.0201 & 0.1937 & 0.0795 & 0.1363 & 8.0\% & 7.3\% \\
\textbf{w/o prompt} & 0.0261 & 0.2941 & 0.0657 & 0.1376 & 9.3\% & 8.5\% & 0.0217 & 0.2075 & 0.0806 & 0.1408 & 8.8\% & 7.8\% \\
\scriptsize{} & \scriptsize{(-5.7\%)} & \scriptsize{(-2.4\%)} & \scriptsize{(-5.3\%)} & \scriptsize{(-10.0\%)} & \scriptsize{(-12.0\%)} & \scriptsize{(-13.3\%)} & \scriptsize{(-8.0\%)} & \scriptsize{(-7.1\%)} & \scriptsize{(-1.4\%)} & \scriptsize{(-3.3\%)} & \scriptsize{(-10.0\%)} & \scriptsize{(-6.8\%)} \\
w/o alignment & 0.0279 & 0.3075 & 0.0680 & 0.1491 & 9.8\% & 8.6\% & 0.0228 & 0.2135 & 0.0823 & 0.1457 & 9.3\% & 8.2\% \\
\scriptsize{} & \scriptsize{(-13.0\%)} & \scriptsize{(-7.0\%)} & \scriptsize{(-9.0\%)} & \scriptsize{(-19.2\%)} & \scriptsize{(-18.1\%)} & \scriptsize{(-14.7\%)} & \scriptsize{(-13.4\%)} & \scriptsize{(-10.2\%)} & \scriptsize{(-3.5\%)} & \scriptsize{(-6.9\%)} & \scriptsize{(-16.3\%)} & \scriptsize{(-12.3\%)} \\
w/o meta learning & 0.0254 & 0.2902 & 0.0642 & 0.1309 & 8.6\% & 7.8\% & 0.0210 & 0.2018 & 0.0799 & 0.1385 & 8.2\% & 7.5\% \\
\scriptsize{} & \scriptsize{(-2.8\%)} & \scriptsize{(-1.0\%)} & \scriptsize{(-2.9\%)} & \scriptsize{(-4.6\%)} & \scriptsize{(-3.6\%)} & \scriptsize{(-4.0\%)} & \scriptsize{(-4.5\%)} & \scriptsize{(-4.2\%)} & \scriptsize{(-0.5\%)} & \scriptsize{(-1.6\%)} & \scriptsize{(-2.5\%)} & \scriptsize{(-2.7\%)} \\
\bottomrule
\end{tabular}}
\label{tab:ablation_phoenix_houston}
\end{table*}

To evaluate the contributions of different components in UniDisMob, we conduct comprehensive ablation studies across four cities—Guilin, Boston, Phoenix, and Houston. We successively ablated three modules: (1) w/o prompt: removes the physics-informed prompt module. (2) w/o alignment: removes the physics-guided alignment module. (3) w/o meta learning: removes the meta-learning mechanism and uses a unified model without city-specific adaptation.

The results are shown in Tables~\ref{tab:guilin_boston} and~\ref{tab:ablation_phoenix_houston}. Removing the physics-guided alignment module has the most significant impact on performance, especially on metrics that describe disaster-induced mobility disruptions, such as Decay Rate and Decay Speed. This highlights the critical role of explicitly aligning spatiotemporal decay patterns under disasters with individual trajectories to ensure physical consistency in the generated mobility patterns. 
Removing the physics-informed prompt module leads to moderate performance drops across multiple metrics, indicating that the physics-informed prompt effectively guides the model to perceive disaster conditions and learn mobility patterns that better reflect real-world disaster scenarios.
Removing the meta-learning mechanism results in relatively smaller performance degradation, but consistently negative effects are observed across all metrics and cities. This confirms the importance of meta-learning in adapting to the heterogeneity across different cities.